# Gate-controlled low carrier density superconductors:

# Toward the two-dimensional BCS-BEC crossover

**Authors**

Y. Nakagawa,[1] Y. Saito,[1] T. Nojima,[2] K. Inumaru,[3] S. Yamanaka,[3] Y. Kasahara,[4] Y. Iwasa[1,5]*

**Affiliations**

[1] *Quantum-Phase Electronics Center (QPEC) and Department of Applied Physics, The University of Tokyo, Tokyo 113-8656, Japan.*

[2] *Institute for Materials Research, Tohoku University, Sendai 980-8577, Japan.*

[3] *Department of Applied Chemistry, Hiroshima University, Higashi-Hiroshima 739-8527, Japan.*

[4] *Department of Physics, Kyoto University, Kyoto 606-8502, Japan.*

[5] *RIKEN Center for Emergent Matter Science (CEMS), Wako 351-0198, Japan.*

* iwasa@ap.t.u-tokyo.ac.jp

**Abstract**

Superconductivity mostly appears in high carrier density systems and sometimes exhibits common phase diagrams in which the critical temperature $T_c$ continuously develops with carrier density. Superconductivity enhanced in lightly doped regime has seldom been reported, although it is an ideal direction towards the crossover between Bardeen–Cooper–Schrieffer (BCS) and Bose–Einstein condensate (BEC) limits, where the behavior of Cooper pairs changes dramatically. Here we report transport properties and superconducting gaps in single-crystalline lithium-intercalated layered nitrides ($Li_xHfNCl$ and $Li_xZrNCl$) down to the low-doping regime, enabled by a combination of ionic gating device and tunneling spectroscopy. Upon the reduction of doping, both systems display the increase of $T_c$, up to 24.9 K especially in $Li_xHfNCl$, with the concomitant enhancement of two-dimensionality leading to a pseudogap state below 35.5 K as well as the increase of superconducting coupling strengths $2\Delta/k_BT_c$ reaching 5.9. Such behavior in the carrier density region as low as $10^{20}$ cm$^{-3}$ indicates that lightly doped two-dimensional superconductors exhibit the unprecedented nature that is distinct from that in conventional superconductors, and offer a new route to access BCS-BEC crossover.

MAIN TEXT

# I. INTRODUCTION

Superconductivity at very low carrier density is one of the most fascinating phenomena in condensed matter physics. The signature of anomalous electron pairing far above $T_c$ has recently been detected by a single-electron transistor configuration in SrTiO$_3$ [1]. For another example, FeSe, which is a superconducting semimetal in bulk, shows the magnetic-field-induced transition between two superconducting states [2]. These features are in stark contrast to those in conventional superconductors described by the Bardeen-Cooper-Schrieffer (BCS) theory, and attributed to their small Fermi energies $E_F$, which are comparable to their superconducting gap $\Delta$. However, the universal pairing mechanism in such a crossover regime of weak-coupling BCS and strong-coupling Bose-Einstein condensate (BEC) limits, where $\Delta/E_F \sim 0.5$ [3], still remains to be resolved. Thus, it is highly desired to explore materials exhibiting superconductivity with low carrier densities as well as to develop a new tool for continuous change of carrier density down to an ultralow level.

Candidate materials that we propose are layered nitrides $MNX$ ($M$ = Ti, Zr, Hf; $X$ = Cl, Br, I), which are semiconductors and exhibit superconductivity by the intercalation of various metals into their van der Waals gap (Fig. 1(a)) [4-6]. As summarized in Figure S1 in Supplementary Section I [7], $T_c$ of both Li intercalated HfNCl and ZrNCl is anomalously high among superconductors with carrier densities lower than $10^{22}$ cm$^{-3}$. Furthermore, in the Zr system, $T_c$ increases from 11 to 15 K with the reduction of carrier density [20]. This unique superconducting phase diagram that differs from the dome-shaped ones [21,22], as well as the extremely small $\gamma$ value determined by the specific heat measurements [23], raised arguments on potential mechanisms [24-27]. However, such an enhancement has not been reported yet in the Hf system, where we can expect that the higher $T_c$ bring the system closer to the BCS-BEC crossover point than in the case of the Zr system. The obstacle in the previous experiments on layered nitrides is that all of them have been performed only in polycrystalline samples. The grain boundaries and the deficiencies enhance the electron localizations and prevent us from clarifying intrinsic properties especially at the low carrier density limit. Therefore, for further discussion on the pairing properties, the study of single crystal with continuous change of carrier density is crucial.

Here, we demonstrate the precise control of electrochemical intercalation processes in an electric-double-layer transistor (EDLT), employing micro-scaled single crystals of layered nitrides $M$NCl, fabricated by the mechanical exfoliation. We have designed a device structure that allows us *in situ* measurements of resistivity and tunneling conductance during the electrochemical intercalation process into the bulk preventing surface doping via EDL, and also enables us to access superconductivity in the low carrier density region which cannot be achieved in polycrystalline bulk samples. In this paper, we investigated superconductivity induced by Li interaction in single crystals of both ZrNCl and HfNCl, and find unprecedented superconducting properties, which include concomitant increase of $T_c$, anisotropy, and coupling strength with reducing the carrier density by controlling the amount of intercalated Li.

The device structure was modified from the conventional EDLT devices [22,28,29] so that the surface carrier doping via EDL is prohibited and only bulk carrier doping via intercalation [30,31] is allowed. Figures 1(b) and (c) display an optical image and schematic diagram of the device used in the present paper, respectively. Because the transport channels of the devices are covered by poly(methyl methacrylate) (PMMA), EDL between the top surface of the channel and the electrolyte, LiClO$_4$/PEG (polyethylene glycol), cannot be formed while the Li intercalation process proceeds through the open area of both ends of the single crystal. The device was fabricated in a N$_2$ filled glove box, which is directly connected to an evaporator to avoid exposure to air. Figure 1(d) shows a typical operation of HfNCl and ZrNCl devices at 330 K. Above a certain critical gate voltage, a sharp decrease in resistance was observed, showing a signature of electron doping through Li intercalation [30,31]. After the decrease of resistance is saturated, we cooled down the device and measured the Hall effect at 150 K, which is far below the melting temperature of electrolyte (see the Appendix). We assumed that the obtained carrier density is equal to the amount of intercalated Li.

## II. RESULTS

Figure 2(a) shows the temperature dependence of resistivity for Li$_x$HfNCl and Li$_x$ZrNCl single crystals below 40 K. $T_c$ of Li$_x$ZrNCl, defined as the mid-point of the resistive transition, increases from 11.5 to 15.4 K on the reduction of doping from $x =$ 0.28 to 0.02 (Top panel), being consistent with the bulk measurements on polycrystalline Li$_x$ZrNCl [20]. Unexpectedly, a similar enhancement of $T_c$ was found in Li$_x$HfNCl, in which $T_c$ increased from 19.4 to 24.9 K by reducing the doping $x$ from 0.28 to 0.03. This value is close to the highest $T_c$ of 25.5 K in layered nitrides, which was observed in cointercalated Li$_x$(THF)$_y$HfNCl ($x = 0.48$) [5]. The enhanced $T_c$ in the Hf analog is in stark contrast to the results of polycrystalline samples, in which below $x = 0.10$, superconductivity was absent [32]. The limit of superconductivity in polycrystalline Li$_x$HfNCl is likely due to the electron localization by disorder, and the presently introduced device using single-crystalline samples pushed down the localization limit, enabling us to approach the unprecedented lightly doped regime in Li$_x$HfNCl.

The $\rho(T)$ curves in Fig. 2(a) show a gradual decrease above $T_c$. This feature is attributed to the superconducting fluctuation. Importantly, as seen in both panels of Fig. 2(a), the contribution of this fluctuation seemingly depends on $x$. To quantify the observed excess conductivity, we introduced the Lawrence-Doniach model [33], which describes thermal fluctuation in layered superconductors by considering each layer as a two-dimensional (2D) superconducting sheet interacting with neighbors by the Josephson coupling. In this model, anisotropic parameter $r = 4\xi_z^2/s^2$ is used to describe the strength of interlayer coupling, where $\xi_z$ and $s$ are the out-of-plane coherence length at zero temperature and the interlayer spacing, respectively. Smaller $r$ corresponds to weaker coupling and stronger anisotropy, therefore in the limit of $r \to 0$, the expression of the fluctuation in this model becomes identical to that in the two-dimensional superconductor. For instance, Aslamazov-Larkin contribution is described as

$$\Delta\sigma_{AL} = \frac{e^2}{16\hbar s} \frac{1}{\sqrt{\varepsilon(\varepsilon + r)}} \qquad (1)$$

which shows 2D-like $\varepsilon^{-1}$ behavior in $r \to 0$, where $\varepsilon$ is the reduced temperature $\ln(T/T_c)$. We performed fitting by using $r$ as a fitting parameter (see Supplementary Section II [7]), the obtained $r$ is plotted in Fig. 2(b) as a function of carrier density both for $Li_x ZrNCl$ and for $Li_x HfNCl$. $r$ systematically decreased with reduction of carrier density, reaching 0.0004 for $x = 0.03$ in $Li_x HfNCl$, which indicates that the superconductivity is getting highly 2D. If one compares these results with those for high-$T_c$ cuprates, such as less anisotropic YBCO ($r = 0.1$) [34] or highly anisotropic Bi-2212 ($r = 0.004$) [35], obtained $r$ values for the Zr and Hf systems stand out particularly in the low doping limit. Below $x = 0.05$, superconductivity is regarded as practically 2D. Such a doping-induced dimensional crossover, as well as the highly 2D nature in the low doping limit, is an unprecedented property in layered superconductors, which became clear by introducing the *in situ* measurements of intercalations processes on single crystals.

This doping-induced dimensional crossover was also confirmed by the in-plane $H_{c2}$ measurement of $Li_x ZrNCl$, shown in Fig. 2(c). Here $H_{c2}$ was defined as the magnetic field showing a half value of the normal state resistance. In the lightly doped regime ($x = 0.06$), the temperature dependence of $H_{c2}$ shows square root behavior, which is described by the 2D Tinkham model, whereas the linear behavior in the heavily doped regime ($x = 0.28$) is described by the Ginzburg-Landau model for anisotropic three-dimensional superconductors [36]. Even in highly doped $Li_xZrNCl$ ($x = 0.28$), the ratio with the out-of-plane $H_{c2}$ is still as high as 38. This anisotropy parameter is much larger than that obtained in the compressed pellet of 4.5 for $ZrNCl_{0.7}$ [37], again indicating crucial importance of using single crystals.

Figure 2(d) displays a $T_c$ versus doping level phase diagram for $Li_xHfNCl$ and $Li_xZrNCl$. Gray squares and triangles represent the data reported on polycrystalline powders [20,32]. Red circles and blue diamonds are the data obtained in the present experiments. The Zr system shows fair agreement between the polycrystal and single crystal samples, whereas the Hf analogue indicates that the *in situ* measurement in the single crystal system is able to reach the low carrier density region where $T_c$ is enhanced by nearly 25%.

For elucidating the superconducting coupling strength, we measured tunneling spectra of $Li_xHfNCl$. Figure 3(a) depicts a schematic energy diagram at the interface between the Ti/Au electrode and $Li_xHfNCl$. Because the mother compound HfNCl is an insulator before doping, we found that the Schottky barrier remains even after Li intercalation, and works as a tunnel junction [Fig. 3(a)] [38]. By using the formula for the depletion layer [39] (see Supplementary Section VIII [7]), the Schottky barrier thickness in $Li_xHfNCl$ was estimated to be 0.5 and 0.2 nm for $x = 0.04$ and 0.28, respectively. These values are comparable to the conducting Hf-N layer (0.2 nm), and the effective thickness of the barrier may be enhanced by the insulating Cl layer and van der Waals gap. Such moderate barrier enables us to use electrodes both for tunnel junctions and for the voltage probes in the four-terminal transport measurement. The right side of Fig. 3(a) shows an optical micrograph and the measurement configuration for tunneling spectroscopy. We applied dc and small ac voltage (or current) and measured between the branch of the narrow electrode and a Hall bar. Because the Schottky barrier behaves as a tunneling barrier, dc bias $V$ dependence of differential conductance is described as

$$\frac{dI}{dV}(V) = A\int_{-\infty}^{\infty} N(E)\left(-\frac{\partial f(E+eV)}{\partial V}\right)dE \qquad (2)$$

where $A$ is a constant, while $N(E)$ and $f(E)$ are the density of state and the Fermi distribution function, respectively. Figure 3(b) shows the tunneling spectra at 2 K for $Li_x$HfNCl with $x = 0.04$ and 0.28, clearly demonstrating the formation of the superconducting gap $\Delta$ in $N(E)$ and its variation depending on $T_c$. Importantly, the data were taken on the same contact of the same sample in different intercalation processes.

Figure 3(c) shows temperature variation of the tunneling spectra for $Li_x$HfNCl ($x = 0.04$) after subtraction of the channel resistance and normalization by the spectrum at 50 K (see Supplementary Section III [7]). An interesting feature is that the gap is not completely closed even at 30 K, which is a higher temperature than $T_c$. The red line in Fig. 3(d) shows the temperature dependence of the normalized zero bias conductance (ZBC) $\Delta G/G_N$, where $\Delta G = G - G_N$, $G = dI/dV$ ($V$=0), and $G_N$ is $G$ under an out-of-plane magnetic field of 9 T. The suppression of ZBC above $T_c$ provides another piece of evidence for the existence of the pseudogap state. The gap opening temperature $T^*$ was determined to be 35.5 K where ZBC is suppressed by 2%. The excess conductivity shown with the blue line in Fig. 3(d) has the onset temperature proximate to $T^*$, indicating that the observed pseudogap state between $T^*$ and $T_c$ is due to the superconducting fluctuation. The pseudogap-like behavior observed in the nuclear magnetic resonance measurement of the lightly doped $Li_x$ZrNCl ($x = 0.08$) [40] polycrystals likely has the same origin, being different from the electron correlation driven or magnetic correlation driven pseudogap in high-$T_c$ cuprates [21]. We note that the pseudogap is observed also in highly doped Li$_x$HfNCl ($x = 0.28$), and $T^*$ is 24 K, which is relatively close to $T_c$ (see Supplementary Section IV and Fig. 4(a), inset [7]). The temperature range of the pseudogap state is doping-dependent.

By fitting the tunneling spectra using Eq. (2) where $N(E)$ is the Dynes function [41], we determined the gap value $\Delta$ at each temperature and plotted in Fig. 4(a) for Li$_x$HfNCl ($x = 0.04$ and 0.28) (see Supplementary Section VI [7]). Since the gap exists far above $T_c$, the temperature dependence of $\Delta$ cannot be explained by the BCS theory, in which $\Delta$ arises at $T_c$ approximately according to a function of $\sqrt{1-T/T_c}$. The gap opening temperature $T^*$ in the inset of Fig. 4(a) is derived from the midpoint between the lowest zero-gap temperature and the highest finite-gap temperature. $T^* = 35.5$ and 23 K at $x = 0.04$ and 0.28, respectively, are in good agreement with $T^*$ obtained from ZBC. We calculated doping dependence of $2\Delta/k_BT_c$ at 2 K [Fig. 4(b)], which is the barometer of the coupling strength. Li$_x$HfNCl is a member of the strong-coupling superconductor having gap ratios larger than 3.5 in the BCS theory, being consistent with the result of break-junction tunneling spectroscopy in polycrystalline Li$_{0.5}$(THF)$_y$HfNCl [42] and HfNCl$_{0.7}$ [43] samples. For the Zr analog, we plotted the data from the specific heat measurement on polycrystalline samples [44]. In both systems, the ratio increases with reducing the doping level $x$. Figure 4(c) shows a summary of $T_c$ as a function of $2\Delta/k_BT_c$ both for Li$_x$HfNCl and Li$_x$ZrNCl systems. A common trend in layered nitrides indicates that the coupling strength well controls the increase of $T_c$ in this system.

### III. DISCUSSION

The gap opening below $T^*$ is a very common trend in 2D systems, which is observed in metallic films [45] and LaAlO$_3$/SrTiO$_3$ interfaces [46]. In the latter system, $T^*$ was found to increase with decreasing the carrier density, showing the same carrier density dependence as that in the present system. However, layered nitrides exhibit the opposite $T_c$-$x$ relation, where both $T_c$ and $T^*$ increases with decreasing the carrier density $x$ as seen in the inset of Fig. 4(a). This indicates that an alternative picture can be argued.

An important clue is that the superconducting coupling strength is dramatically enhanced in the low density region. Figure 5 shows the ratio of the superconducting gap and the Fermi energy, $\Delta/E_F$, as a function of $x$. Both Zr and Hf systems display an increase of $\Delta/E_F$ by order of magnitude with decreasing $x$. Particularly in Hf analogs, $\Delta/E_F$ reaches 0.12 at $x = 0.02$. This is a remarkable enhancement, which is caused by the combined effect of the reduction of carrier density and the concomitant increase of $\Delta$. The increase of $T_c$ with the increase of $\Delta/E_F$ as well as increased $T^*$ is reminiscent of the general phase diagram of BCS-BEC crossover, when the coupling strength is increased from the BCS limit. It is also noted that the pseudogap region between $T_c$ and $T^*$ is wider in $x = 0.04$ than in $x = 0.28$. Taking these results into account, we conclude that our system is approaching the BCS-BEC crossover region near $\Delta/E_F \sim 0.5$ [3] by reducing the carrier density with the ionic gating technique.

Recently, the BCS-BEC crossover has been discussed in bulk single crystal FeSe [2] and magic-angle twisted-bilayer graphene [47], where $\Delta/E_F = 1$ or $T_c/T_F = 0.08$ is observed, respectively. In the latter, the gate control of carrier density is playing a crucial role in a similar manner to the present result. The present system, Li$_x$HfNCl and Li$_x$ZrNCl, is of particular interest because approaching the BCS-BEC crossover occurs at rather high $T_c$, and thus may play an important role in the study of BCS-BEC crossover. Regarding the specific pairing mechanism, the present results give a strong constraint to theories. Among many theoretical proposals on superconductivity in Li intercalated HfNCl and ZrNCl systems [24,26,27], the recently reported one taking the valley polarization fluctuation [26,27] at $K$ and $K'$ points into account seems interesting, because it explains the increase of $T_c$ with reducing the carrier density in 2D multivalley superconductors. It is also noted that the experimentally found gap-opening temperature $T^*$ is close to the predicted $T_c$ of 40 K [27] in Li$_x$HfNCl, indicating that the theoretically predicted $T_c$ might correspond to the temperature at which the superconducting fluctuation sets in. Such an enhancement is predicted to be universal in 2D doped semiconductors. Not only the layered nitrides, but also other 2D materials like transition-metal dichalcogenides can be an attractive platform to investigate the BCS-BEC crossover.

To conclude, the voltage control of electrochemical intercalation process using EDLT devices revealed that superconducting Li$_x$HfNCl and Li$_x$ZrNCl are potentially a model system to approach the 2D BCS-BEC crossover, in which an increase of $T_c$ and novel exotic aspects of superconductivity are anticipated. We emphasize that gate-controlled nanodevices potentially offer opportunities to access a new state of matter, which is difficult to reach in bulk systems.

## ACKNOWLEDGMENTS

We thank M. Nakano and M. Yoshida for discussions and experimental support. This work was supported by JSPS KAKENHI Grant Numbers JP25000003, JP17J08941, JP15J07681. Y.N. was supported by Materials Education program for the future leaders in Research, Industry, and Technology (MERIT).

# APPENDIX: MATERIALS AND METHODS

## A. Device fabrication

Pristine ZrNCl and HfNCl single crystals were grown by a chemical vapor transport method using vacuum-sealed silica glass tubes and a two-step heating method [48]. They were mechanically exfoliated into thin flakes with tens of nanometers in thickness using the Scotch-tape method, and transferred onto Si/SiO$_2$ substrates. Ti (5 nm)/Au (70 nm) electrodes were deposited on an isolated thin flake, as well as the gate electrode on the substrate, by the standard electron beam lithography process. We covered the device with PMMA resist and developed the outer area of the channel. Lithium electrolytes were prepared by dissolving LiClO$_4$ in PEG. The melting temperatures of PEG used in our experiments, $M$ = 600 and 100 000, are 15 and 65 °C, respectively. We added ethanol for easier handling in case of PEG ($M$ = 100 000). After putting a droplet of the electrolyte solution on the flake and the gate electrode, samples were dried in vacuum (lower pressure than $10^{-4}$ Torr) for 1 hour to eliminate residual moisture, oxygen, and ethanol.

## B. Transport measurements

The temperature-dependent resistance was measured with a standard four-probe geometry in a Quantum Design Physical Property Measurement System with two kinds of lock-in amplifiers (Stanford Research Systems Model SR830 DSP and Signal Recovery Model 5210). The gate voltage was applied by a Keithley 2400 source meter at 300 or 330 K under high vacuum. After cooling down to low temperature, the chamber was purged with He. A horizontal rotator probe was used to measure angular dependence of the upper critical field.

### C. Tunneling spectroscopy

To input ac and dc excitations, a multifunction generator (NF Corporation WF1974) or ac and dc current source (Keithley Model 6221) were used and both setup output identical signals. Voltages were probed with an ac lock-in amplifier and a dc voltmeter (Keithley Model 2182A) in the former setup, and with a dc voltmeter (Keithley Model 2182A, differential conductance mode) in the latter setup.

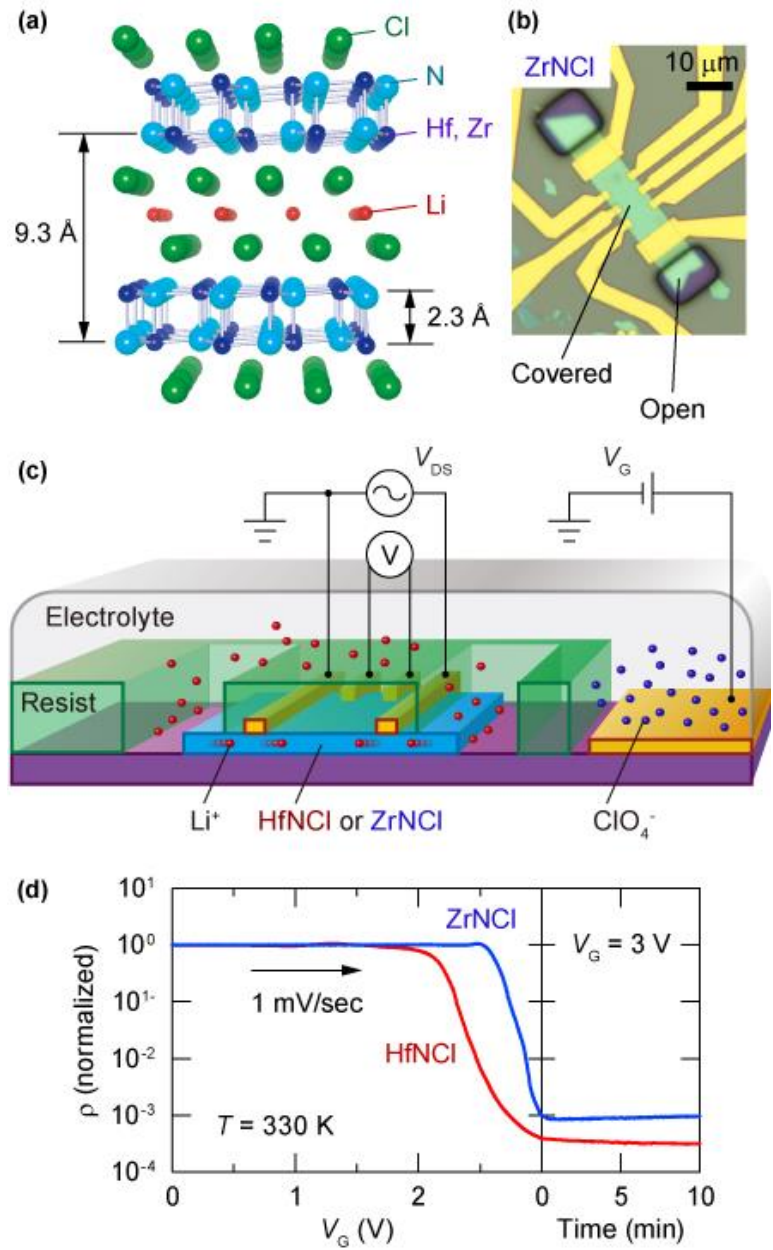

**FIG. 1. Ionic-gated intercalation in layered nitrides. (a)** Crystal structure of HfNCl (ZrNCl). Hf (Zr)-N double honeycomb layer and trigonal Cl layers are stacked along *c* axis. Intercalated Li ions provide electrons to Hf (Zr)-N layers. **(b)** Optical microscope image of a flake of a layered nitride on a $SiO_2$/Si substrate patterned into a Hall bar configuration. It is covered with PMMA resist except for the area outside the channel to prevent electrolyte from touching the HfNCl (ZrNCl) surface to form the conducting channel. **(c)** Schematic of intercalation driven by a positive gate voltage. Li ions were intercalated to the crystal, and the resistance was *in situ* monitored with a small source-drain voltage. **(d)** Resistivity of ZrNCl (blue) and HfNCl (red) during intercalation at 330 K followed by waiting at $V_G$ = 3 V to make it homogeneous. The resistivity decreases because of electron doping.

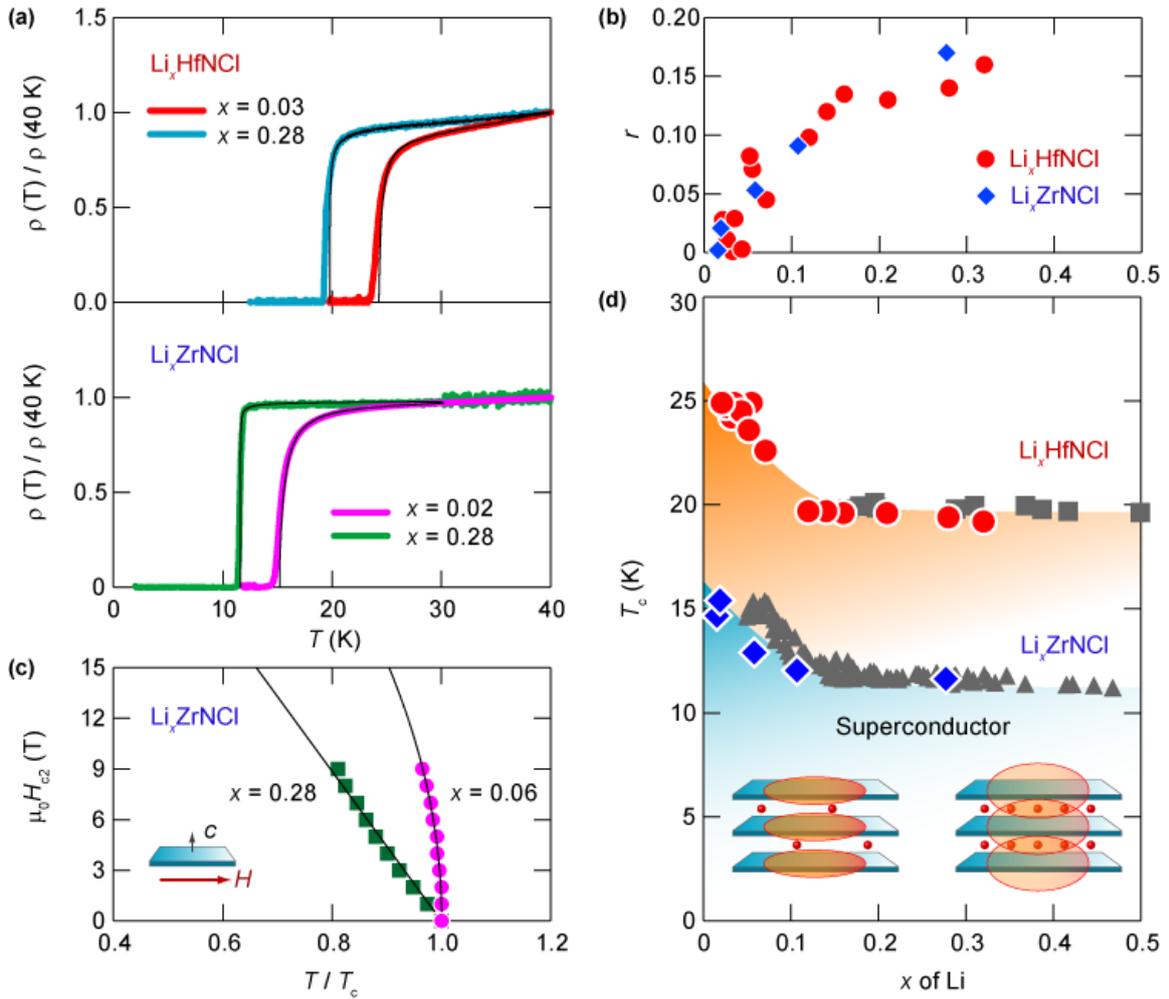

**FIG. 2. Doping dependence of $T_c$ and dimensionality.** (a) Superconducting transition in Li-doped HfNCl (Top) and ZrNCl (Bottom). Lower doped samples ($x = 0.02, 0.03$) show higher critical temperatures $T_c$ with larger superconducting fluctuation than higher doped samples ($x = 0.28$). Fits to excess conductivity described by the Lawrence-Doniach model are drawn with solid black curves. (b) Anisotropic parameter in the Lawrence-Doniach model as a function of doping level. Smaller values at light doping level both in Li$_x$HfNCl (red) and in Li$_x$ZrNCl (blue) correspond to stronger two-dimensionality. (c) In-plane upper critical field $H_{c2}$ as a function of normalized temperature with $T_c$. Difference of initial rise indicates the change in dimensionality on the reduction of doping from $x = 0.28$ to $0.06$. (d) Phase diagram of layered nitrides. Red circles and blue diamonds indicate $T_c$ in single crystals of Li$_x$HfNCl and Li$_x$ZrNCl, respectively, whereas gray squares and triangles correspond to the polycrystalline samples of Li$_x$HfNCl and Li$_x$ZrNCl, respectively [20,32]. The common feature is the enhancement of $T_c$ in the lightly doped regime, where superconductivity shows two-dimensionality.

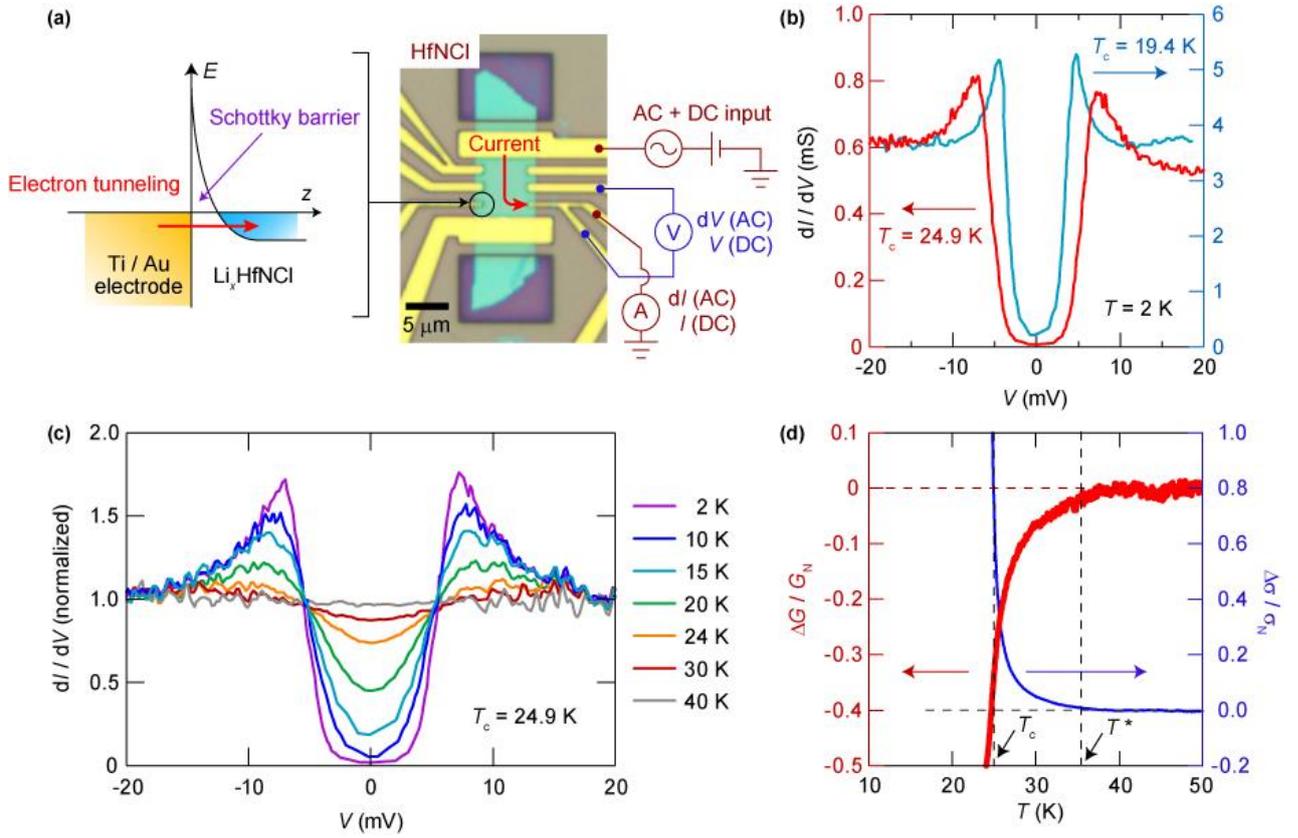

**FIG. 3. Tunneling spectroscopy in Li$_x$HfNCl. (a)** (left) Schematic band diagram at the junction between electrode and Li$_x$HfNCl. Electrons tunnel through the Schottky barrier. (right) Illustration of measurement setup. Small ac voltage superimposed on dc voltage is applied and the voltage drop at the junction below a thin electrode branch is measured. Channel resistance can be measured in the same cool down process. **(b)** Raw data of tunneling spectra in different doping regimes. The different values of the superconducting gap are easily observable, reflecting different $T_c$ of 24.9 (red) and 19.4 K (blue). Difference of the conductance in the normal state appearing at high-biases indicates the different thickness of the Schottky barrier. **(c)** Tunneling spectra at 2, 10, 15, 20, 24, 30, and 40 K. Upon warming, the gap gets smaller but still exists at 30 K, which is a higher temperature than $T_c$. The spectra are normalized by using the spectrum at 50 K as a back ground. **(d)** Temperature dependence of the zero-bias conductance, or $dI/dV$ ($V = 0$). The gap starts opening at 35.5 K, which is almost identical to the rising point of excess conductivity, drawn with blue curves.

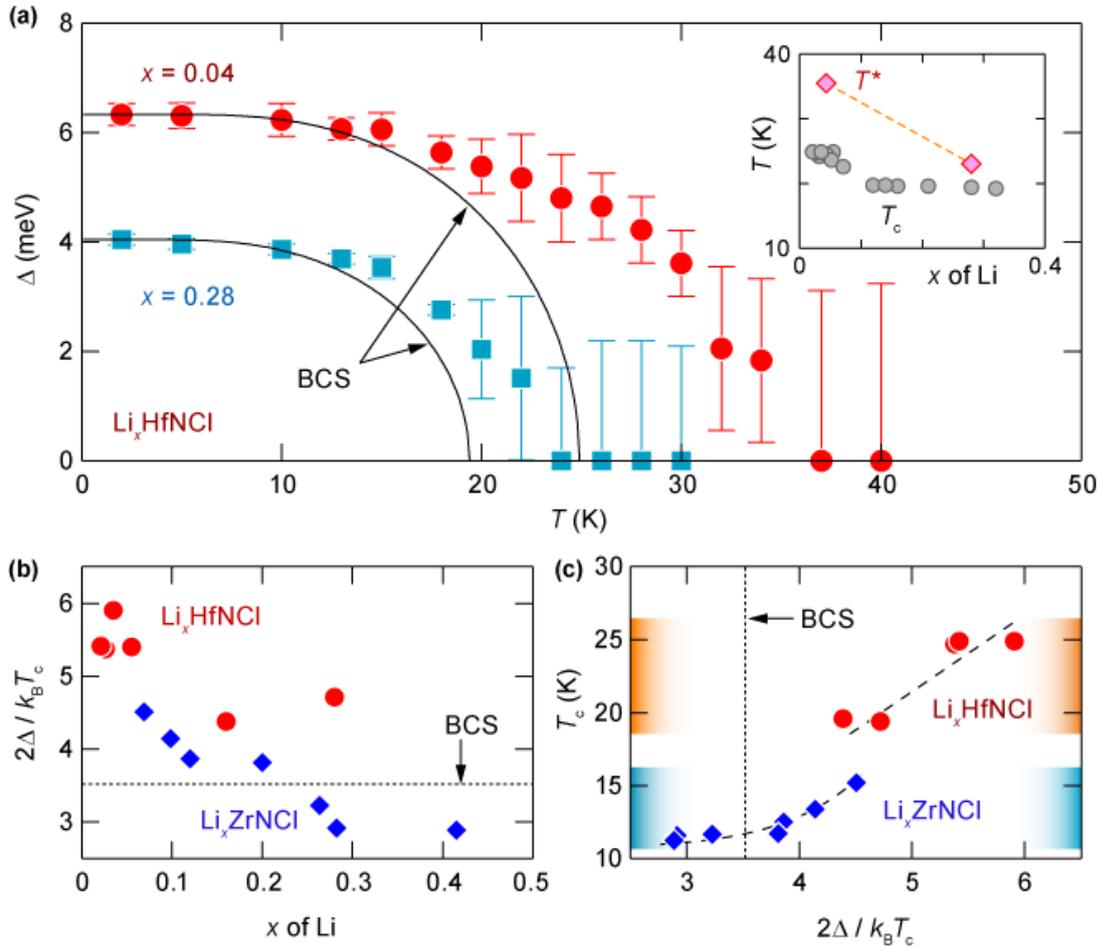

**FIG. 4. Superconducting gap and gap ratio in layered nitrides (a)** Temperature dependence of $\Delta$ for $x = 0.04$ (red circles) and 0.28 (blue squares). Black curves are fits to the BCS-shape function, where $T_c$ was determined by the resistive transition. Error bars represent uncertainties in the fitting (see Supplementary Section VI [7]). The inset shows the doping dependence of $T^*$ and $T_c$. **(b)** Doping dependence of gap ratio $2\Delta/k_B T_c$. Red circles indicate the data single crystals of Li$_x$HfNCl. The data for polycrystalline Li$_x$ZrNCl (blue diamonds) are taken from the result of specific-heat measurement [44]. They increased upon the reduction of doping, or the enhancement of two-dimensionality, getting beyond the BCS weak-coupling value of 3.53 (dotted line). **(c)** Critical temperature $T_c$ as a function of $2\Delta/k_B T_c$, or coupling strength. The $T_c$ increases monotonically up to 24.9 K. The data in Li$_x$ZrNCl are taken from the specific-heat measurement [44]. The dotted line and dashed line are the BCS prediction (3.53) and guide to the eye, respectively.

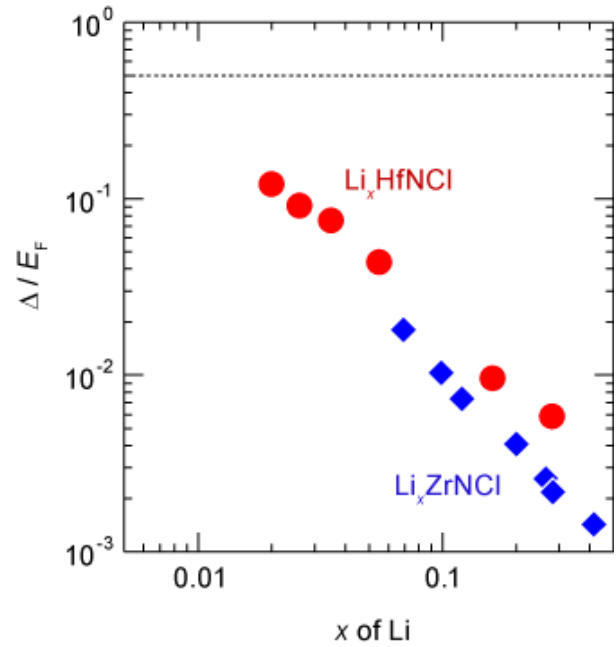

**FIG. 5. Doping dependence of the ratio of superconducting gap to Fermi energy.** The value of $\Delta/E_F$ increases with decreasing the doping level approaching $\Delta/E_F = 0.5$ (dotted line), which indicates the BCS-BEC crossover region [3]. Red circles indicate the present data in $Li_xHfNCl$, whereas the data in polycrystalline $Li_xZrNCl$ (blue diamonds) are derived from Ref. [44].

Supplementary Materials for

# Gate-controlled low carrier density superconductors:

# Toward the two-dimensional BCS-BEC crossover

Y. Nakagawa, Y. Saito, T. Nojima, K. Inumaru, S. Yamanaka,

Y. Kasahara, and Y. Iwasa*

* iwasa@ap.t.u-tokyo.ac.jp

**Contents**

I. **Comparison with other low-carrier-density superconductors**

Layered nitrides ($Li_xHfNCl$ and $Li_xZrNCl$) show the enhancement of superconducting critical temperature $T_c$ by reducing the carrier density $n_{3D}$ as in the main text (Fig. 2(d) in the main text). This peculiar behavior becomes apparent when we plot the $T_c$-$n_{3D}$ relationship together with other low-carrier-density superconductors in Fig. S1. The data for $Li_xHfNCl$ was first obtained in this work. $n_{3D}$ was calculated from the Hall coefficient for $Li_xHfNCl$, $Li_xZrNCl$, $SrTiO_3$ [8], GeTe [9], and PbTe [10], and $Cu_xBi_2Se_3$ [11], whereas $n_{3D}$ of $La_{2-x}Sr_xCuO_4$ and diamond was extracted from dopant density [12, 13]. For FeSe, both electron and hole densities were derived from Shubnikov–de Haas oscillations [14].

$SrTiO_3$ [8] is a well-known low-carrier-density superconductor, which shows a dome-shaped phase diagram by electron doping. The $T_c$ of $SrTiO_3$ increases with decreasing carrier density until the peak where $n_{3D} = 1 \times 10^{20}$ cm$^{-3}$, and then $T_c$ starts to decrease. GeTe [9], PbTe [10], and diamond [13] also exhibit superconductivity with small carrier densities, and their $T_c$ monotonically decreases with reducing carrier densities. Therefore, $T_c$ is limited at low-carrier-density regime in other superconductors. In stark contrast, $T_c$ of layered nitrides are prominently high among low-carrier-density superconductors and keep increasing down to $n_{3D} = 5 \times 10^{20}$ cm$^{-3}$, where superconductivity in $La_{2-x}Sr_xCuO_4$ [12] disappears. Such peculiar behavior of $Li_xHfNCl$ and $Li_xZrNCl$ suggests that layered nitrides have a potential to reach the lower carrier density regime with sustaining their relatively high $T_c$, and thus the Cooper pairs in those systems may show unique nature as investigated in FeSe [2].

## II. Analysis of excess conductivity

To analyze excess conductivity due to superconducting fluctuation above $T_c$, we took three terms into account.

$$\Delta\sigma = \Delta\sigma_{AL} + \Delta\sigma_{MT} + \Delta\sigma_{DOS} \qquad (S1)$$

They describe the contribution from Aslamazov-Larkin (AL) process, Maki-Thompson (MT) process, and reduction of density of state (DOS), respectively. The specific expression of each term is given based on Lawrence-Doniach model, which describes excess conductivity of a layered superconductor with variable interlayer coupling [33].

As mentioned in the main text, the AL term is described as

$$\Delta\sigma_{AL} = \frac{e^2}{16\hbar s}\frac{1}{\sqrt{\varepsilon(\varepsilon+r)}} \qquad (S2)$$

where $s$ is the interlayer distance, $\varepsilon$ is the reduced temperature $\ln(T/T_c)$ and $r$ is the anisotropic parameter. The MT term is summation of two terms, regular (reg) MT process and anomalous (an) MT process.

$$\Delta\sigma_{MT(reg)} = -\frac{e^2}{2\hbar s}\kappa_1 \ln\left(\frac{2}{\varepsilon^{1/2}+(\varepsilon+r)^{1/2}}\right) \qquad (S3)$$

$$\Delta\sigma_{MT(an)} = \frac{e^2}{4\hbar s(\varepsilon-\gamma_\phi)}\ln\left(\frac{\varepsilon^{1/2}+(\varepsilon+r)^{1/2}}{\gamma_\phi^{1/2}+(\gamma_\phi+r)^{1/2}}\right) \qquad (S4)$$

where $\kappa_1$ and $\gamma_\phi$ are the functions of temperature $T$ and other parameters. By assuming that $k_B T\tau/\hbar$ is much smaller than 1, where $\tau$ is the electron scattering time, $\kappa_1$ becomes constant value of 0.691, and $\gamma_\phi$ can be regarded as a linear function of $1/T$ [15]. The DOS term acts as a negative contribution,

$$\Delta\sigma_{DOS} = -\frac{e^2}{2\hbar s}\kappa_2 \ln\left(\frac{2}{\varepsilon^{1/2}+(\varepsilon+r)^{1/2}}\right) \qquad (S5)$$

where $\kappa_2$ is a similar function to $\kappa_1$, which becomes constant value of 0.691 when $k_B T \tau / \hbar \ll 1$.

For the analysis, we used the conductivity under out-of-plane magnetic fields of 9 T as the normal state conductivity $\sigma_N$. The excess conductivity $\Delta\sigma = \sigma - \sigma_N$ was fitted by the models described in equation (S1) with temperature-independent offsets [16] to make $\Delta\sigma = 0$ at 40 K. The distance between centers of $[ZrN]_2$ or $[HfN]_2$ layers at $x = 0$ [20, 32] was used for $s$. We assumed that $\kappa_1 = \kappa_2 = 0.691$ and $\gamma_\phi = a/T$, where the value of the parameter $a$ is between 0.5 and 5. The typical fitting result are shown in Fig. S2. We note that the fitting only with the AL term can reproduce the doping dependence of $r$ (Fig. 2(c) in the main text).

### III. Subtraction of the channel resistance from the tunneling spectra

In our configuration for the tunneling spectroscopy, the channel resistance $R_{ch}$ makes the non-negligible contribution to the differential conductance $dI/dV$ at the temperature above $T_c$. Before normalization of the tunneling spectra, we subtracted the contribution of the channel according to the following procedure.

As shown in Fig. S3(a), the system is described as a series circuit, which consists of the tunneling junction and the channel. The measured differential resistance $dV/dI$ (the inverse of differential conductance) becomes sum of the differential channel resistance and differential tunneling resistance, and expressed as

$$\frac{dV}{dI}(V) = \frac{dV_j}{dI}(V_j) + \frac{dV_{ch}}{dI}(V_{ch}) = \frac{dV_j}{dI}(V_j) + R_{ch} \quad (S6)$$

where $V = V_j + V_{ch}$ is the measured voltage, and $V_j$ and $V_{ch}$ are the voltage drop at the junction and the channel, respectively. $R_{ch}$ is assumed to be ohmic resistance which is independent of the bias voltage. The additional resistance $R_{ch}$ makes the measured differential conductance smaller than the differential conductance of the junction, which reflects the density of state (DOS) of the sample, whereas the measured bias voltage $V$ becomes larger than the voltage drop at the junction $V_j$.

In Fig. S3(b), we plotted the raw tunneling spectra measured with the DC-biased AC excitation at different temperatures. The DC bias current $I_{DC}$ ranges from -10 to 10 μA. Here we focus on the $dI/dV$ at high biases around 20 mV, which corresponds to the DOS of the normal state, and is expected to be independent of the temperature when $R_{ch} = 0$. The spectra at 2 and 20 K show good agreement at high biases around 20 mV, whereas the spectra at higher temperatures shows the suppression of the conductance depending on the temperature. This deviation clearly originates from the existence of $R_{ch}$, as expressed in equation (S6). (However, the existence of the gap above $T_c$ is visible even in the raw data.)

According to equation (S6), we can obtain the differential resistance of the channel by plotting $dV/dI - R_{ch}$ as the function of $V - V_{ch} = V - R_{ch} \times I_{DC}$. Although we can measure $R_{ch}$ by usual 4-terminal configuration, $R_{ch}$ should be different in the tunneling measurement because current path is different (expressed as $R_{ch}^{4T}$ and $R_{ch}^{tun}$, respectively). Therefore, we estimated $R_{ch}^{tun}$ at 50 K from the $dV/dI$ spectra, and determined the ratio $R_{ch}^{tun}/R_{ch}^{4T}$ at 50 K. The $R_{ch}^{tun}$ at $T < 50$ K was calculated by multiplying $R_{ch}^{4T}$ at $T$ by $R_{ch}^{tun}/R_{ch}^{4T}$ at 50 K. The latter factor is independent of the $dV/dI$ spectra. As represented in Fig. S3(c), $R_{ch}^{tun} = 600$ Ω is the best estimation at 50 K, because the corrected spectra agree with the spectra at 2 K in high bias regime, as expected from the discussion above. We applied this factor ($R_{ch}^{tun}/R_{ch}^{4T} = 4.3$) at 50 K to all the spectra at lower temperatures, and obtained the pure tunneling spectra.

The corrected $dI/dV$ spectra are plotted in Fig. S3(d). For 2 and 20 K, the spectra are identical to that in Fig. S3(b), because $R_{ch}$ is definitely zero. By dividing the conductance at 2-40 K by that at 50 K, normalized tunneling spectra are obtained as shown in Fig. 3(c) in the main text.

## IV. Doping dependence of the tunneling spectra

Figures S4(a) and (b) shows the normalized tunneling spectra at various temperatures in Li$_x$HfNCl where $x$ = 0.28 and 0.04 ($T_c$ = 19.4 and 24.9 K), respectively. (Figure S4(b) is the same figure as Fig. 3(c) in the main text.) They were measured in the same junction of the same sample, but in different doping level. The gap structure disappeared in heavily doped regime at 30 K, but is clearly recognized at the same temperature in lightly doped regime.

## V. Contribution of the fluctuation on the zero-bias conductance

We observed the decrease of the zero-bias conductance above $T_c$ as shown in the Fig. 3(d) in the main text. It can be ascribed to the reduction of the density of state $N(E)$, i.e. the opening of the superconducting gap, due to the superconducting fluctuation. The pseudogap state originating from superconducting fluctuation was also reported in TiN thin films, which is a conventional 2D superconductor [45].

In the 2D system, the reduction of $N(E)$ affects the zero-bias tunneling conductance as follows [33]:

$$\frac{\Delta G}{G_N}(\varepsilon) \cong 2Gi \ln \varepsilon \qquad (S7)$$

where $G_N$ is the conductance in the normal state, $\Delta G$ is the reduction of the conductance from $G_N$, $\varepsilon$ is the reduced temperature $\ln T/T_c$, and $Gi$ is the Ginzburg-Levanyuk parameter.

The characteristic behavior of equation (S3) is the linear behavior in the semi-logarithmic plot as the function of $\varepsilon = \ln T/T_c$. Figure S5 shows such plot of the zero-bias conductance, which is the same data as in Fig. 3(d) in the main text. The linear behavior appeared, and $Gi$ is determined to be 0.047, half of the value 0.1 for TiN. Because the equation (S3) is valid for $\varepsilon \gg Gi$ [33], the non-linear behavior below $\varepsilon \sim$

## VI. Fitting of tunneling spectra

After normalization of tunneling spectra, we performed fitting to obtain the value of superconducting gap $\Delta$. We used equation (2) in the main text as the fitting function for the tunneling conductance, where $A$ is determined by the normalization process in Supplementary Section III. Also, $N(E)$ is Dynes function [41]:

$$N(E) = \mathrm{Re}\left( \frac{E - i\Gamma}{\sqrt{(E - i\Gamma)^2 - \Delta^2}} \right) \quad (S8)$$

where $\Gamma$ is a measure of the decay rate of the quasiparticle. The temperature dependence of the fitting parameters, $\Delta$ and $\Gamma$, is plotted in Fig. S6a) and (b) for Li$_x$HfNCl ($x = 0.28$ and $0.04$), respectively. Corresponding tunneling spectra with fitting curves are summarized in Fig. S6(c) and (d). The value of $\Delta$ in Fig. S6(a) corresponds to the Fig. 4(a) in the main text, and normalized spectra in Fig. S6(c) and (d) corresponds to the spectra in Fig. 3(c) in the main text and Fig. S4(a), respectively. The peak structures were smeared at high temperatures, leading to the large uncertainties of $\Gamma$ and $\Delta$. The length of error bars is determined as the difference between the qualitative fit (plotted in figures) and the numerical fit.

## VII. Effect of the resist cover

To confirm that EDLT effect is removed by the resist cover, we fabricated a device of ZrNCl which contains two areas of channels covered and uncovered with the E-beam resist, as shown in Fig. S8(a). Figure S8(b) shows the resistances in each area during gating at 300 K. While the resistance of the covered area did not respond to the gate voltage $V_G$, the resistance of the open area decreases showing usual EDLT operation [28]. This measurement proves that the formation of EDL doesn't occur through the resist and the observed superconductivity is induced by intercalation of Li into bulk. Furthermore, the result of the covered channel also indicate that intercalation does not occur by covering the edge of the sample.

VIII. **Estimation of the thickness of the tunneling barrier**

We estimated the thickness of the Schottky barrier $W$, which acts as the tunneling barrier, by using the formula [39]:

$$W = \sqrt{\frac{2\varepsilon\phi_b}{eN}} \quad (S9)$$

where $\varepsilon$ is the dielectric constant, $\phi_b$ is the built-in potential, $e$ is the elementary charge, and $N$ is the carrier density.

We assumed $\varepsilon = 5\varepsilon_0$ in Li$_x$HfNCl, where $\varepsilon_0$ is the dielectric constant of vacuum, according to the static dielectric constant in Li$_x$ZrNCl [17, 18]. We calculated $\phi_b$ as $\phi_b = \phi_m - \phi$, where $\phi_m$ is the work function of adhesion layer Ti, 4.1 V, and $\phi$ is the bottom of the conduction band measured from vacuum level. For $\phi$, we used a theoretically estimated value 3.76 eV on a monolayer HfNCl [19]. With these assumptions, W was calculated as 0.5 and 0.2 nm for $x = 0.04$ and 0.28, respectively.

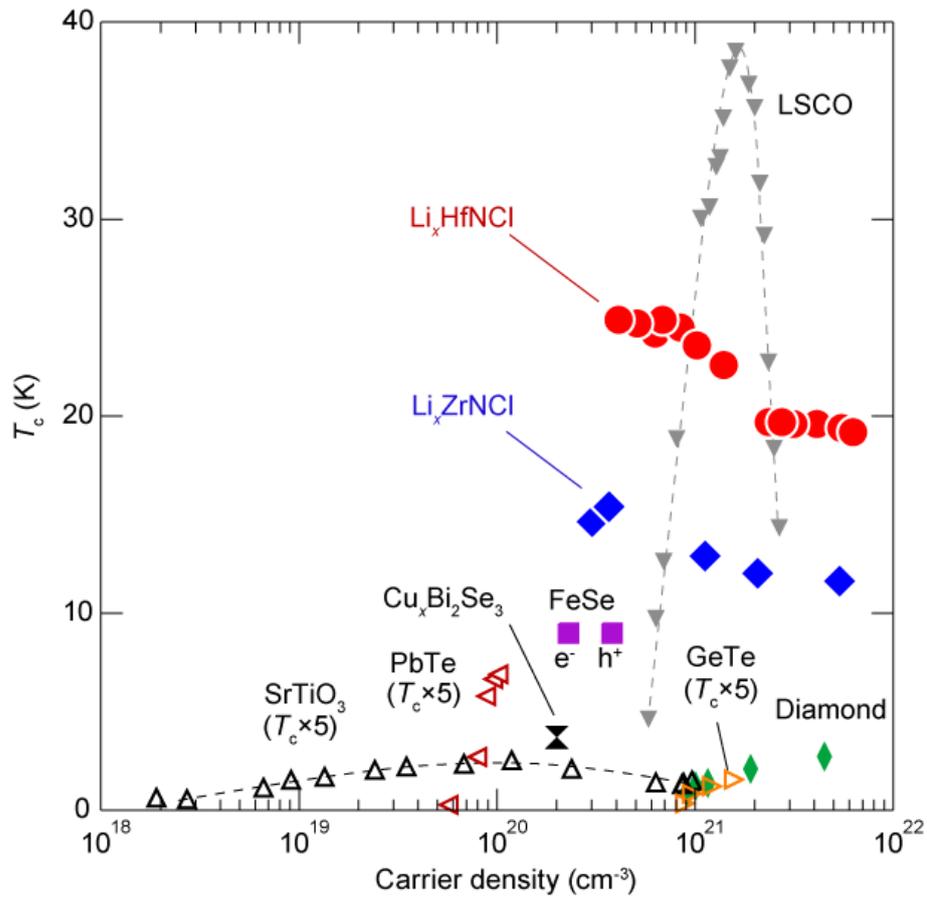

**FIG. S1. $T_c$ as a function of three-dimensional charge carrier density in low-carrier-density superconductors.** $T_c$ of SrTiO$_3$ [8], GeTe [9], and PbTe [10] (open symbols) are multiplied by 5. Carrier density of La$_{2-x}$Sr$_x$CuO$_4$ [12] and diamond [13] corresponds to the dopant density. For semimetallic FeSe [14], both electron and hole densities are plotted. Dashed lines are guide to the eye.

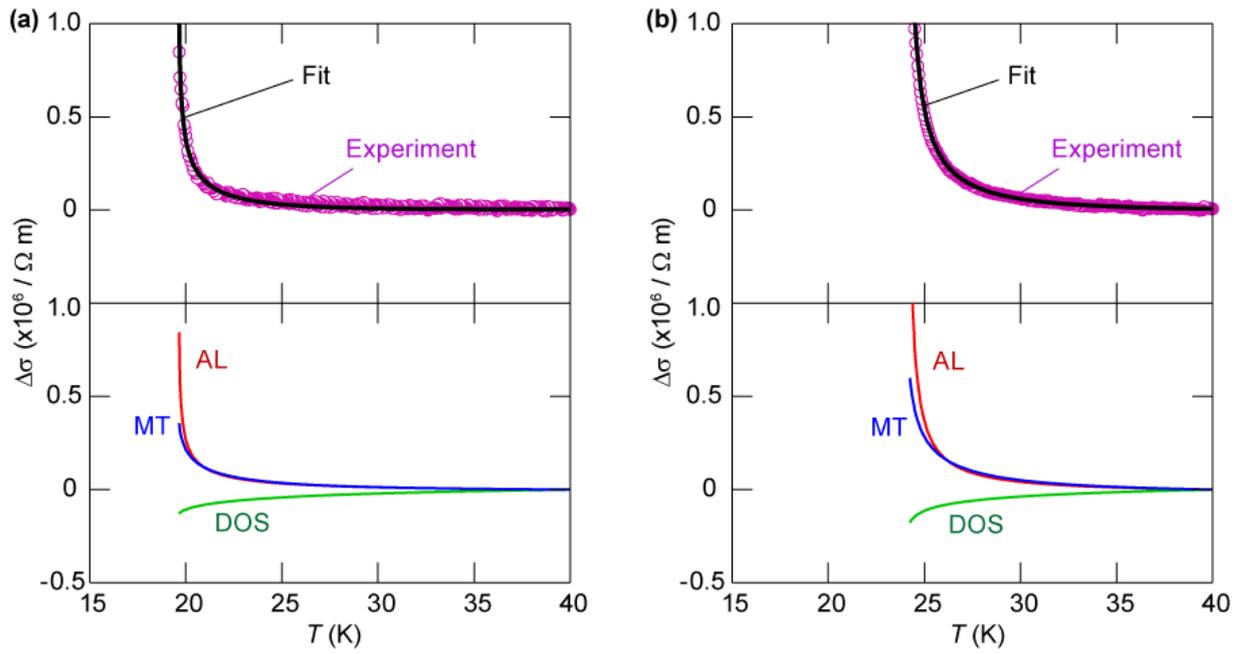

**FIG. S2. Fitting of excess conductivity**. (Top) The experimental data and fitting results for $x = 0.03$ (**a**) and 0.28 (**b**) in Li$_x$HfNCl. (Bottom) The contributions of AL term (red), MT term (blue), and DOS term (green) are plotted separately.

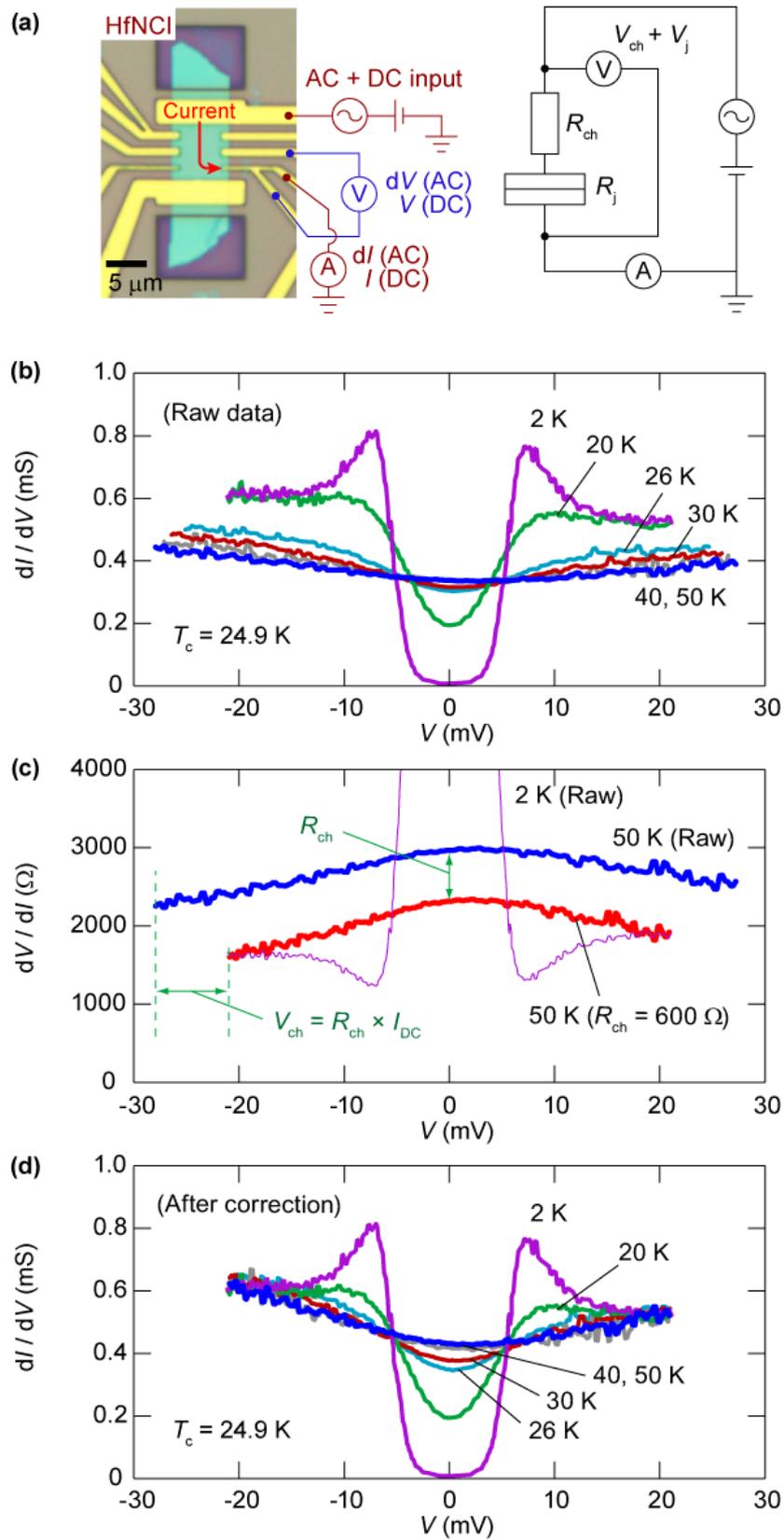

**FIG. S3. Normalization process of tunneling spectra.** (a) (left) Illustration of measurement setup, same as the right figure of Fig. 3(a). (right) The equivalent circuit of the measurement setup. The channel resistance $R_{ch}$ is included in series with the resistance of the tunneling junction $R_j$. (b) Raw data obtained by tunneling

spectroscopy at various temperature (2, 20, 26, 30, 40 and 50 K). The differential conductance at high bias voltage are constant at low temperatures, and decrease due to channel resistance above $T_c$. **(c)** The d$I$/d$V$ spectra at 2 and 50 K. The channel resistance $R_{ch}$ is estimated to be 600 Ω to eliminate the discrepancy at high biases. **(d)** The spectra after the subtraction of channel resistance, showing the voltage drop at tunnel junction only.

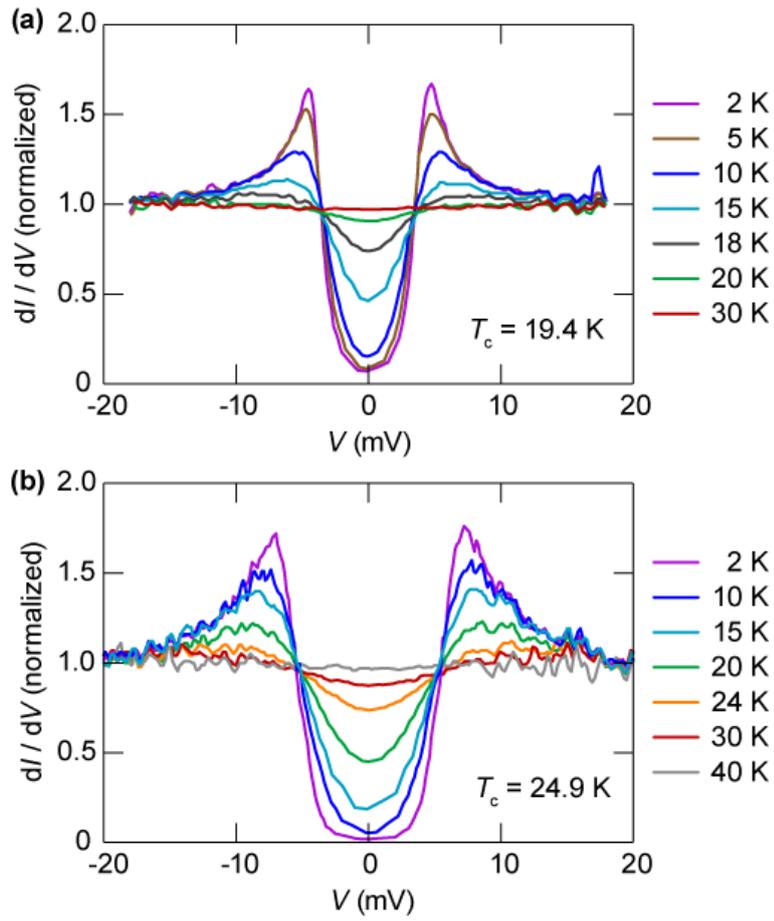

**FIG. S4. Tunneling spectra in different doping level. (a)** Tunneling spectra in heavily doped regime, where the $T_c$ is low. The gap structure is absent at 30 K. **(b)** Tunneling spectra in lightly doped regime, showing a signature of the gap at 30 K.

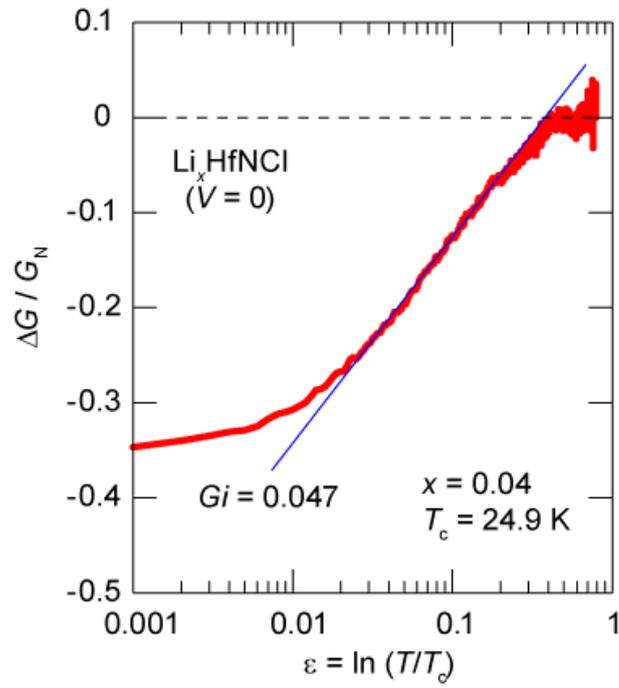

**FIG. S5. Analysis of zero-bias conductance.** The linear behavior as the function of log $\varepsilon$ represents the reduction of DOS in the 2D superconductor.

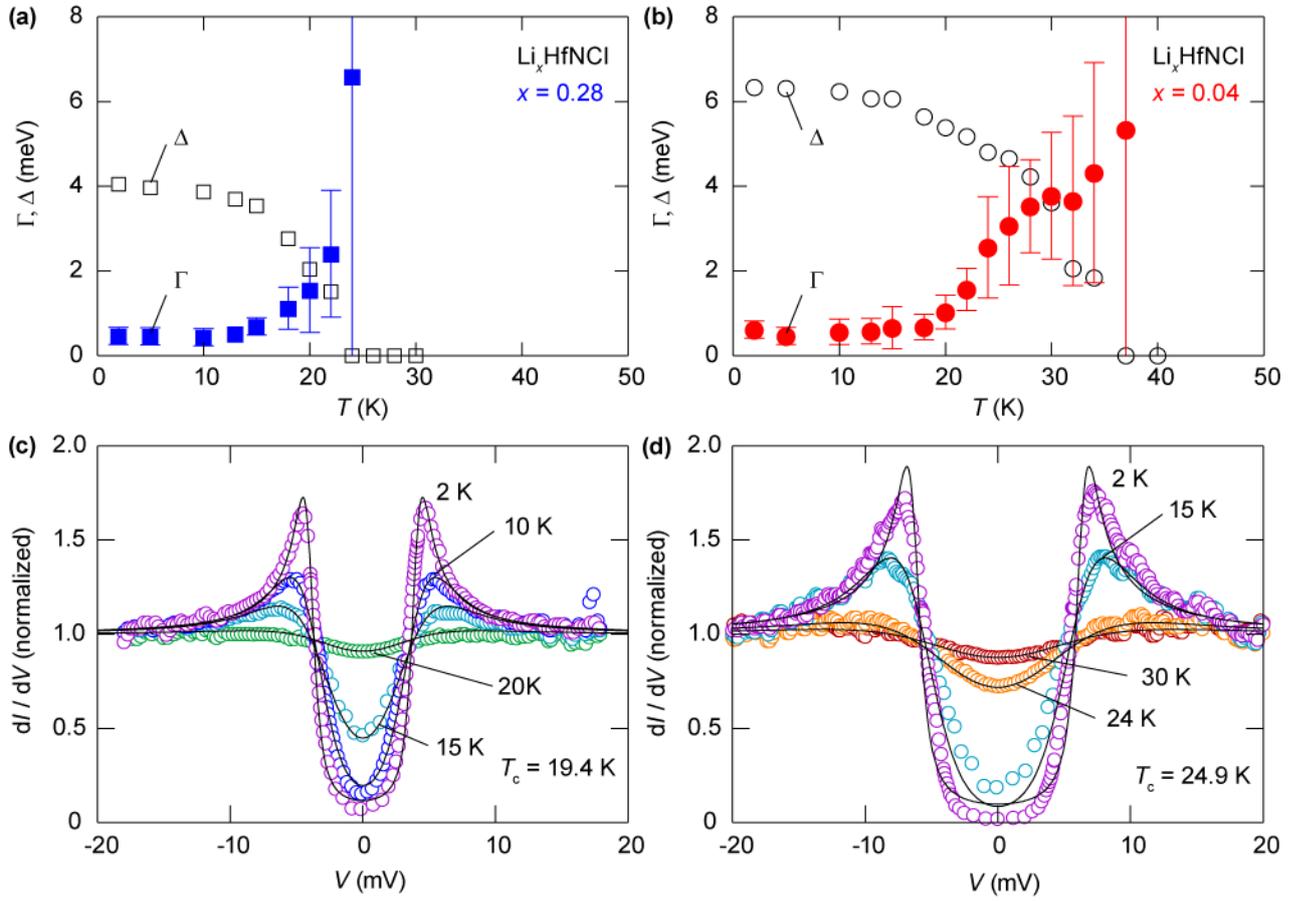

**FIG. S6. Fitting results for tunneling spectra.** (**a**) and (**b**) Temperature dependence of $\Gamma$ (filled symbols) and $\Delta$ (open symbols) at $x = 0.28$ (**a**) and 0.04 (**b**). Error bars represent uncertainty in the fitting (see the text). Error bars for $\Delta$ are plotted in Fig. 4(a) in the main text. (**c**) and (**d**) The tunneling spectra at $x = 0.28$ (**c**) and 0.03 (**d**) with fitting curves using Dynes function.

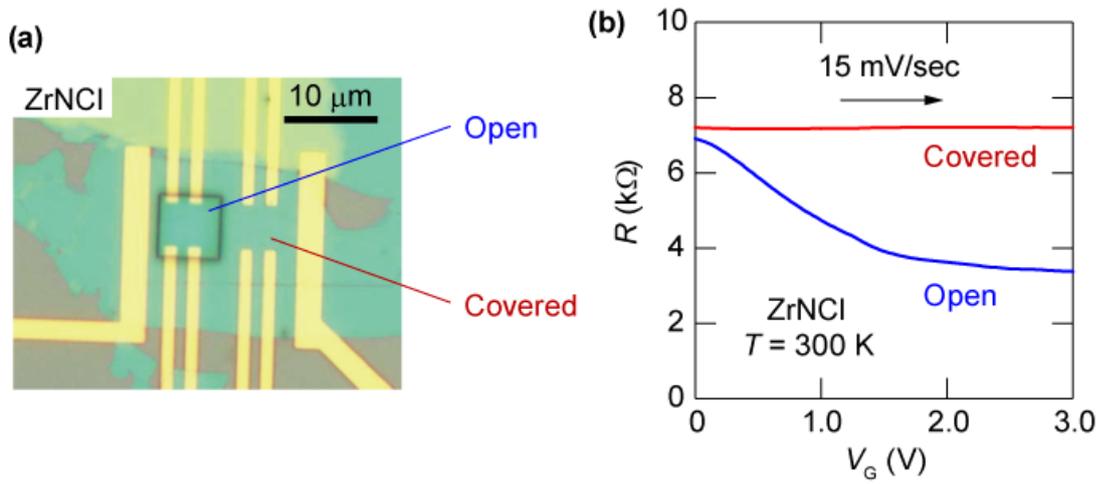

**FIG. S7. Confirmation of the resist cover.** (a) Optical microscope image of a device which has open area and covered area with PMMA resist. (b) The gating response of open area (blue) and covered area (red). The resistance of open area decreases due to the formation of EDL.